\documentclass[12pt,preprint]{aastex}
\usepackage[dvips]{color} 
\usepackage{amssymb}

\newcommand{\myemail}{mizuta@post.kek.jp}

\slugcomment{revised to ApJ}

\shorttitle{Thermal Radiation from GRB Jets}
\shortauthors{Mizuta et al.}

\begin{document}

\title{Thermal Radiation from GRB Jets}

\author{Akira Mizuta\altaffilmark{1,2},
Shigehiro Nagataki\altaffilmark{3},
and Junichi Aoi\altaffilmark{3}}

\altaffiltext{1}{Theory Center, Institute of Particle and Nuclear Studies, KEK (High Energy Accelerator Research Organization), 1-1 Oho, Tsukuba 305-0801, Japan}
\altaffiltext{2}{Center for Frontier Science, Chiba University
Yayoi-cho 1-33, Inage-ku, Chiba, 263-8522, Japan}
\altaffiltext{3}{Yukawa Institute for Theoretical Physics,
Kyoto University, Kitashirakawa Oiwake-cho, Sakyo-ku, Kyoto, 606-8502, Japan}
\email{AM, e-mail: \myemail}

\begin{abstract}
In this study,
the light curves and spectrum of the photospheric thermal radiation
from ultrarelativistic gamma-ray burst (GRB) jets are calculated
using 2D relativistic hydrodynamic simulations of
jets from a collapsar.
As the jet advances, 
the density around the head of the jet decreases,
and its Lorentz factor reaches
as high as 200 at the photosphere
and 400 inside the photosphere.
For an on-axis observer,
the photosphere appears concave shaped
due to the low density and high beaming factor of the jet.
The luminosity varies because of
the abrupt change in the position of the photosphere 
due to the internal structure of the jet.
Comparing our results with 
GRB090902B,
the flux level of the thermal-like component is similar
to our model, although the peak energy looks a little bit
higher (but still within a factor of $2$).
From the comparison, we estimate that the bulk Lorentz factor 
of GRB090902B
is $\Gamma \sim 2.4 \times 10^2 (r/10^{12} \rm cm$)
where $r$ is the radius of the photosphere.
The spectrum for an on-axis observer
is harder than that for an off-axis observer.
There is a time lag of a few seconds for high energy bands
in the light curve.
This may be the reason for the delayed onset of
GeV emission seen in GRB080916C.
The spectrum below the peak energy is a power law and
the index is $2.3 \sim 2.6$
which is softer than that of single temperature plank distribution
but still harder
than that of typical value of observed one.
\end{abstract}


\section{INTRODUCTION}
\label{sec:intro}
Gamma-ray bursts (GRBs), the most energetic explosion in the universe,
show a non-thermal spectrum, implying that GRBs originate from
an optically thin region (e.g. Piran (1999) and references therein).
On the other hand, some long GRBs are associated with a supernova (SN)
explosion (e.g., Woosley and Bloom (2006) and references therein).
Are GRBs already optically thin when they break out the
their progenitors ? The answer is obviously no, since the jet density
should be high and Lorentz factor should be low.
Thus, we can expect to observe the thermal radiation from the photosphere
prior to GRBs.
Even in the prompt phase, there may be a thermal radiation 
component coming from the photosphere.

About 25 years ago,
it was suggested GRBs should show thermal spectra unless they
are coming from relativistically-moving objects~\citep{Goodman86}. 
Similar to our motivation, 
the thermal radiation from the photosphere
of GRBs were analytically discussed in some
papers~(\citet{Blinnikov99}, \citet{ryde05},
\citet{peer07}, \citet{lixin07}, and \citet{peer08}).
In the photospheric model \citet{Lazzati09},
thermal radiation is scattered by relativistic electrons
producing non-thermal photons of GRBs (e.g. Rees and M{\'e}sz{\'a}ros
(1999); Ioka et al. (2007); Toma et al. (2009 and 2010)).
The effects of magnetic
fields, for example the reconnection,
on the emission have been discussed
in some papers~(\citet{giannios06} and \citet{zhang09}).

Thermal components have been observed in some GRBs
from the data obtained from BATSE catalog and/or BeppoSAX
catalog~(\citet{ghirlanda07}, and \citet{ryde09}).
A thermal component has been reported for
GRB060218 associated with SN2006aj~\citep{campana06,liang06}.
Recently, {\it Fermi Gamma-ray Space Telescope} observed GRB090902B,
showing a clear
spectrum consist of two components~\citep{Ryde10} with one appearing
to be a thermal component.

In this study, we investigate the thermal radiation 
from the photosphere of GRBs through numerical
simulations. 
Some studies have been carried out on the jet
propagation of GRBs (e.g. Mizuta \& Aloy (2009) and references
therein); however these studies did not determine the photosphere's location
and the number of thermal photons radiated from GRBs except for Lazzati et al. (2009).
In this study,
we determine the location and shape
of the photosphere of GRBs as a function of
time and estimate the light curve and spectrum of the thermal
radiation.
Here we present thermal spectrum and light curves 
for different viewing angles and light curves for
several energy bins corresponding to {\it Fermi} and {\it Swift} data.
We also discuss the unique  spectrum of GRB090902B and
show the delayed onset of hard photons possibly related to
the one seen in GRB080916C \citep{Abdo09},
and we discuss viewing angle effects 
on short GRBs and X-ray flash (X-ray flare).

\section{NUMERICAL SETUP}
\label{sec:method}
This section describes the progenitor model
and the numerical conditions of the hydrodynamic calculations.
Although the structure of the progenitor at the pre-SN stage is not known,
realistic calculations of stellar evolution
of massive stars have been recently done by \citet{Woosley06},
\citet{Yoon05} and \citet{Yoon06}.
We choose one of models in \citet{Woosley06}.
The model is named 16TI whose core has a relatively high spin at the pre-SN stage;
this progenitor model is also used by
\citet{Morsony07}, \citet{Lazzati09} and \citet{Morsony10}.
At zero age, the progenitor is assumed to have a mass of
16 solar masses
and low metallicity (0.01 $Z_\odot$).
At the pre-SN stage, the total mass is 13.95 solar masses and
the progenitor radius $(r_{*})$ is $4\times 10^{10}$~cm.

The 2D spherical coordinate system ($r\times \theta$) is used for hydrodynamic calculation,
assuming axisymmetry and equatorial plane symmetry.
The computational domain covers the region of
$10^9$~cm $\le r \le3\times 10^{12}$~cm and $0^{\circ}\le\theta\le 90^{\circ}$.
We remap the density profile from the progenitor model
into the computational domain
($r_{\rm min}\le r\le r_{\rm *}$),
assuming spherical symmetry.
The reflective boundary condition is applied
at the polar axis and the equatorial plane.
Since the stellar wind theory is not well known so far and
a weak wind is expected due to very low metallicity,
we assume the gas outside the progenitor is dilute
and the power law distribution $\rho =1.8\times 10^{-14}(r/r_{\rm *})^2
{\rm g~ cm}^{-3}$.
Since this weak wind does not affect to the jet dynamics
and the opacity, the model corresponds to the most luminous case.

The radial grid consists of $N_r=2640$ points,
uniformly spaced in $\log{r}$, extending from $r_{\rm
  min}=10^9$\,cm to $r_{\rm max}=3\times 10^{12}$\,cm.
The smallest radial
grid spacing, at $r_{\rm min}$, is $\Delta r_{\rm min}=10^7$\,cm,
while the largest one, at $r_{\rm max}$, is $\Delta r_{\rm
  max}=7.6\times 10^9$\,cm.
The uniform 120 polar grid is spaced
between $0^{\circ}\le \theta\le 30^{\circ}$,
i.e., $\Delta \theta =0.25^{\circ}$.
The 60 uniform logarithmic grids are spaced in the range of
$30^{\circ}\le \theta \le 90^{\circ}$.

The mechanism of the jet formation near
the central black hole is still under debate
(but see e.g., Nagataki et al. (2007), Nagataki (2009)).
A very hot, relativistic and
temporary constant jet is injected from the inner most
grid, i.e., $r=r_{\rm min}$ with a half opening angle
$10^{\circ}$.
The velocity vector is parallel to the radial unit vector.
The luminosity of the injected jet is $5.5\times 10^{50}$~erg/s.
Lorentz factor ($\Gamma_0$) is 5 and specific internal energy
($\epsilon_0/c^2$) is 80, where $c$ is the speed of light.
This jet has a potential to be accelerated
to a Lorentz factor of more than 500,
applying relativistic
Bernoulli's principle ; $h\Gamma=$ const, along a stream
line ($h_0\Gamma_0=538$ in our case),
where $h$ is specific enthalpy,
if all thermal energy is converted to kinetic energy
without dissipation. 
We follow the jet propagation until the head
of the jet reaches $r=3\times 10^{12}$~cm.

A special relativistic hydrodynamic code developed by
one of the authors (AM \citet{Mizuta04} and \citet{Mizuta06}) is used
for hydrodynamic simulations.
The version with 2nd order accuracy both in time and space is used,
including some
dissipation to prevent numerical oscillation at strong shocks.

\section{RESULTS}
\label{sec:results}

\subsection{Hydrodynamics}
In the early phase of the evolution,
the jet drills through the progenitor envelope, keeping good collimation,
as shown in previous numerical simulations by
\citet{Aloy00}, \citet{Zhang03,Zhang04}, \citet{Mizuta06},
\citet{Morsony07}, \citet{Mizuta09}, and \citet{Lazzati09}.
The forward shock drives
the envelopes to the high pressure and
high temperature.
A cocoon originates from
the jet material because of the reverse shock
at the head of the jet and surrounds the jet.
Since the density of the jet is considerably lower than that of the
stellar envelopes,
a fast backflow and some vortices appear
in the cocoon \citep{Mizuta10}.
Internal shocks in the jet
appear due to interaction between the jet and the high pressure cocoon and backflows
(Komissarov \& Falle (1997,1998) and \citet{Morsony07}).
The jet reaches the progenitor surface at about $t_{\rm lab}=7$s.

\begin{figure}
\begin{center}
\includegraphics[angle=0,scale=1.]{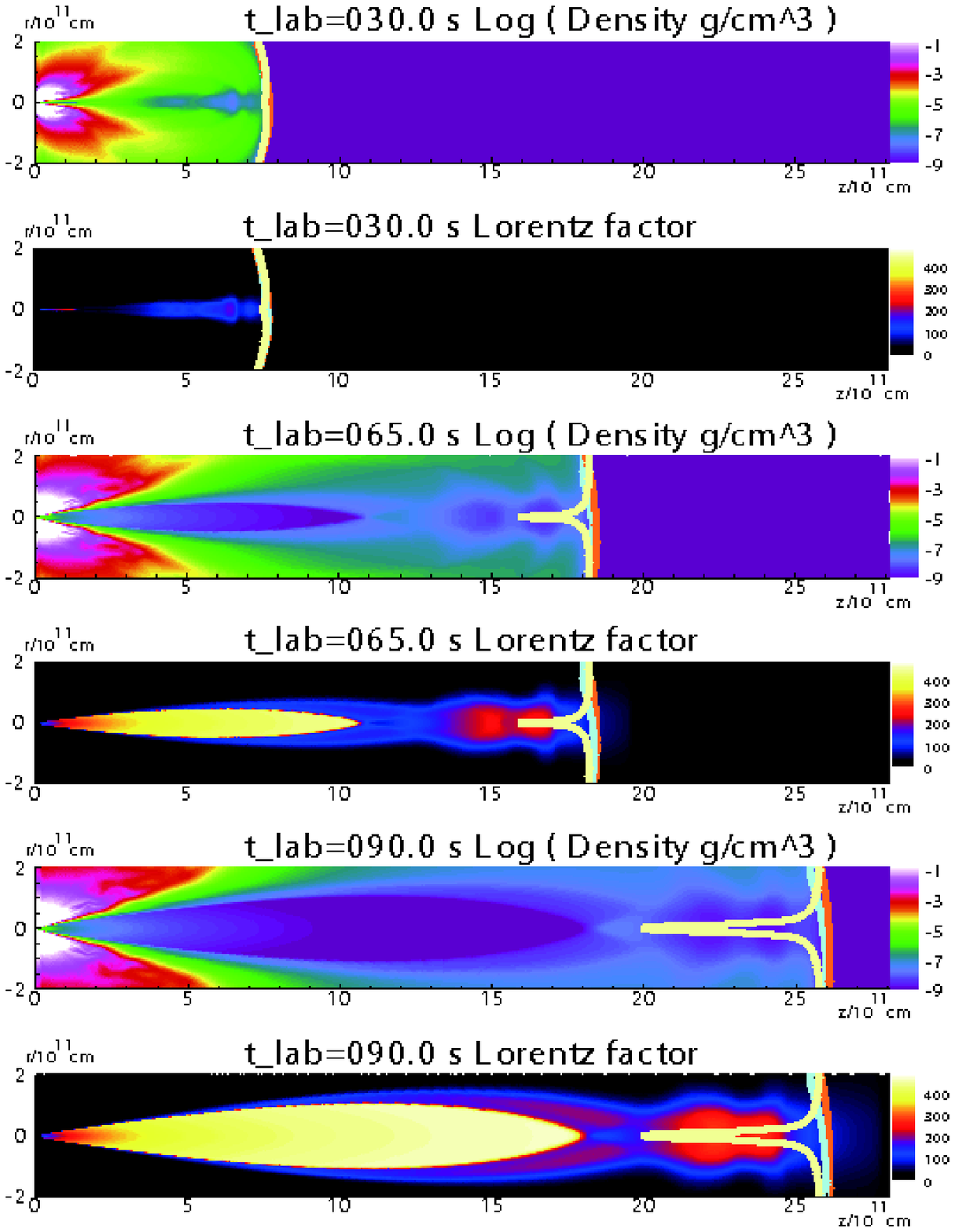}
\end{center}
\caption{The density and Lorentz factor contours
around the jet axis ($z$-axis)
at $t_{\rm lab}=30$ (top two),
65 (middle two) and 90~s (bottom two).
The photospheres for the observer at different viewing angles
(yellow : $\theta=0$, aqua : $\theta=5^{\circ}$, and
orange : $\theta=10^{\circ}$)
are shown.
The observer is at infinity and on $z$-axis ($\theta_{\rm v}=0$)
and at the right-top side ($\theta_{\rm v}>0$).
\label{contours}}
\end{figure}

When the shock breaks out the progenitor surface,
the components near the head of the jet starts expanding,
resulting in a high Lorentz factor component ($\Gamma \sim 200$)
(two top panels of Fig. \ref{contours};
density and Lorentz factor contours at $t_{\rm lab}=30$~s).
The shocked envelopes are also expanding into circumstellar matter.
Since most of the components of the jet are still surrounded by
the expanding cocoon and high density progenitor envelopes,
the jet remains collimated.
The pressure in the cocoon and envelope decreases as they expand.
The components injected from computational boundary 
($r=r_{\rm min}$) at a later phase
can easily propagate in a radial direction without dissipation
(four bottom panels of Fig. \ref{contours}).
The blue regions in density contours and the red and purple regions
in Lorentz factor contours are free expanding regions.
The Lorentz factor exceeds 400.
This free expansion ends at the internal shock.
On the other hand,
since the velocity of the expanding cocoon and stellar envelopes
is less than its sound speed ($\sim 0.5c$),
it is delayed with the head of the jet.
As a result, the head of the jet is quite relaxed,
resulting in a high Lorentz factor component ($\Gamma\sim 200$).
The internal structures imprinted in the jet before the shock break
still remain near the head of the jet.
Such discontinuities can be also seen in a 1D radial plot
(Fig. 2. \cite{Lazzati09}),
but the structure is different from our simulations.
Such difference may be caused by the different code and resolution for
the hydrodynamic calculations.

\subsection{Thermal Radiation from Photosphere}
\label{sec:analysis}
The thermal radiation
from the photosphere is derived in post processing,
from data taken every 1.0 s of the laboratory frame.
In each snapshot of the hydrodynamic simulations,
we find photosphere at unity optical depth 
for the Thomson scattering.
The optical depth ($\tau$) is defined as follows;
\begin{eqnarray}
\label{opacity}
\tau= \int_{x_{\rm ph}}^{\infty}
\sigma_{\rm T}\Gamma(1- \beta \cos \theta )~n~dl,
\end{eqnarray}
where $\sigma_{\rm T}$ is the Thomson scattering cross section,
$\Gamma(\equiv (1-\beta^2)^{-1/2})$ is the Lorentz factor,
$\beta$ is absolute value of velocity normalized by the speed of light,
 $\theta$ is the angle
between the velocity vector ($\vec{\beta}$) and the line-of-sight (LOS),
and $n$ is proper number density of the electron,
$n\equiv 2{\rho/ m_{\rm He}}$, where $m_{\rm He}$ is the mass of helium atom.
Though the progenitor includes many elements except hydrogen,
we assume that all materials consist of
fully ionized helium for simplicity.
The expression in the integral includes the inverse of
beaming factor ($\delta$),
i.e., $\Gamma(1- \beta \cos \theta) (\equiv \delta ^{-1}(\theta))$,
to include the relativistic effect \citep{Abramowicz91}.
Here, we study the effect of the viewing angles,
for an on-axis observer  $\theta_{\rm v}=0$ and 
for off-axis observers,
i.e., the angle between
the jet axis and LOS, $\theta_{\rm v}=
1^{\circ},$ and $2^{\circ}$.

Assuming an observer located at infinity,
the isotropic luminosity of thermal radiation
from the photosphere is evaluated as
\begin{eqnarray}
L_{\rm iso}(\theta_{\rm v})={ac}\int \delta(\theta)^4 T^4 \cos \theta_{\rm ph}dS,
\end{eqnarray}
where $a$ is the radiation constant,
$\theta_{\rm ph}$ is the angle between the LOS and
the normal vector of the emission surface \citep{peer07}.
The isotropic luminosity in the left hand side is considered
by the observer time ($t_{\rm obs}$) which is related with
the laboratory time ($t_{\rm lab}$)
as $t_{\rm obs}=t_{\rm lab}+d/c$, where $d$ is
the distance between each photosphere and the observer.
The local temperature at the photosphere is evaluated
as $T=(3p/a)^{1/4}$, where $p$ is the thermal pressure of the fluid.
The photosphere for $\theta_{\rm v}=0^{\circ}$
on the plane of the jet axis and the observer
is shown by the solid lines in Fig.~\ref{contours}.
Though we also plot the cases for
$\theta_{\rm v}=5^{\circ}$, and $10^{\circ}$,
we do not show light curves and spectrum for these cases due to
too short duration at the observer frame (see, sec 3.3).
The photospheres for each observer almost
overlap at $t_{\rm lab}=30$ s.
As the jet advances, its density decreases.
As a result, the photosphere shifts further inside the jet
(bottom 4 panels in Fig.~\ref{contours} 
at $t_{\rm lab}=65$ and 90~s).
To an on-axis observer, the photosphere, in particular
appears highly concave.

\subsection{Light curves}
\begin{figure}
\begin{center}
\includegraphics[angle=270.,scale=0.33]{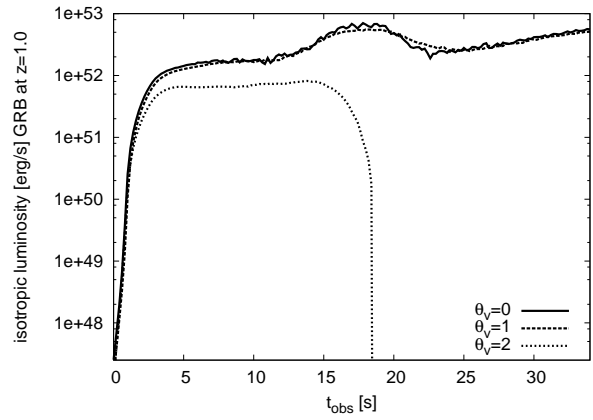}
\caption{The light curves for the observer at different viewing angles,
i.e., $\theta_{\rm v}=0^{\circ} (solid), 1^{\circ}$ (dashed), and
$2^{\circ}$ (dotted).
It is assumed that the burst occurs at the redshift of $z=1$.
The duration of the radiation is longer for the on-axis observers.
The light curve for an on-axis observer is bright and has time
variability.
We should stress the light curves
still
continues and the last part of the light curves for each viewing angle
are not completed.
\label{light1}}
\end{center}
\end{figure}

Hereafter we assume that the burst occurs at the redshift of $z=1$.
Figure \ref{light1} shows the light curves for the observer
at different viewing angles
as a function of the observer time.
The observer detects the radiation at different observer times,
even if the radiation comes from the same laboratory time,
due to the curved photosphere.
We set $t_{\rm obs}=0$ as the burst trigger for each viewing angle.
Though we integrate thermal radiation
only of the whole computational laboratory time ($0<t_{\rm lab}<100$~s)
due to limited CPU time,
the radiation will continue for some cases,
especially on-axis case.
Thus, we should stress that 
the light curves shown here are not completed yet,
though the early phase of the light curve is 
valid.
We need to follow longer timescales
at laboratory time in the future.

The light curves
exhibit time variability,
i.e., for the observers $\theta = 0^{\circ}$ and $1^{\circ}$
second (10 s $<t_{\rm ob}<$ 22 s) and third ($t_{\rm ob}>$ 22 s)
phase after the first phases in the light curve
which exhibits constant luminosity
and continues for more than 32 s.
The feature of light curves is consistent with \citet{Lazzati09}
in which the case of $\theta = 0^{\circ}$ is shown.
Since the observer sees the skin of the jet at the beginning,
the photosphere moves almost at the speed of light.
The duration for the observer is about $1/\gamma^2$ times as long as
that for the laboratory frame, where
$\gamma$ is the Lorentz factor of the motion for the photosphere
($\gamma \sim 10$).
The arrival time of the radiation concentrates within a few seconds
for the observer which causes quick rise.
As time passes, the density near head of the jet
decreases due to the expansion of the jet.
The beaming factor for the on-axis observer is very large,
since the velocity vector is almost parallel to the jet axis.
So, the distance between the forward shock
and photosphere increases, i.e., Eq.\ref{opacity}.
The observer can see the region very deep inside the jet.
The photosphere for the on-axis observer
is concave as shown in bottom four panels of
Fig. \ref{contours}.
Because of discontinuities in the Lorentz factor
and density in the jet,
the photosphere suddenly moves deeper inside the jet.
The second phase caused by the detection of the
internal shock ($\Gamma \sim 200$)
and beaming factor at the photosphere is
more than 400.
The third phase is caused by the radiation from much deeper side.
The Lorentz factor at the photosphere is smaller than
that in second phase.

Figure \ref{flux} shows the photon number flux for
an observer
in the four energy bands as usually used in {\it swift} analysis.
Figure \ref{flux2} is same as Fig. \ref{flux},
but different energy bands as usually used in {\it fermi} analysis.
For $\theta_{\rm v} = 0^{\circ}$, 
only soft photons ($E \le
100$ keV) come in the beginning ($t_{\rm ob} \le$ a couple of seconds)
since the Lorentz factor of the photosphere is still small.
Then, high-energy photons follow.
A few seconds time lag for high energy photons
can be clearly seen in Fig.\ref{flux2} for both
on-axis observer ($\theta_{\rm v}=0^{\circ}$) and off-axis observer
($\theta_{\rm v}=1^{\circ},$ and $2^{\circ}$).

\subsection{Spectrum}
Figure \ref{nufnu} shows the $\nu F_{\nu}$ plot for 
an observer ($\theta_{\rm v}=0^{\circ}, 1^{\circ},$ and $2^{\circ}$).
The spectrum is superposition of Plankian at local rest frame
and each of them is boosted by the beaming factor.
The interval of the whole observer
time is integrated (top panel).
The distribution is power law both below and above the peak energy.
For $\theta_{\rm v} = 0^{\circ}$,
the spectrum has the peak energy at about 270 keV.
A comparison with Fig. \ref{flux2} shows that the spectrum
is soft in the beginning,
but becomes hard later.
Further, as can be expected from Fig.~\ref{flux}
and \ref{flux2}, the spectrum
becomes softer for larger viewing angle cases ($\theta_{\rm v}=1^\circ$ and
$2^\circ$ cases).

The spectrum below the peak energy
has power law index with $2.3 \sim 2.6$ for all viewing angles.
The indices are softer than that of
single temperature plank distribution, see 
two bottom panels in
Fig. \ref{nufnu} in which 
single temperature plank distribution, fitting to the peak energy
and absolute value, is shown by thin dashed lines.
Since the spectrum is superposition of different temperature
Plank distribution,
the power law index is $\sim 2.0$ 
near the peak energy and it continues to one tenth of the peak energy.
The index is still harder than that of typical value of observed one,
i.e., Band function \citep{Band93}.
The index for below the peak energy
band in the spectrum of GRB090902B is about unity \citep{Ryde10}.
If we apply higher angular resolution for hydrodynamic calculation,
the spectrum will consist of much more
different local temperature ($T^{'}$) and beaming factor ($\delta$)
contributions, resulting different observed temperature  ($\delta T^{'}$).
This would make the spectrum softer.
On the other hand,
the index above the peak energy is softer than that of Band function.
This suggests the non-thermal process is necessary to
reproduce the entire GRB spectrum.
We will investigate this non-thermal component in the near future.

\section{DISCUSSION and SUMMARY}
\label{sec:conclusion}
We have calculated the light curves and spectrum of
the photospheric thermal radiation
from GRB jets using 2D relativistic hydrodynamic simulations.
We found that the thermal radiation from the photosphere
of GRB jets is ``observable'' even for cosmologically distant GRBs.
This component can be observed as a precursor or a thermal component in
the prompt phase.
The simulation needs further study to ascertain whether
this component can be observed in the afterglow phase.

As for GRB090902B, the flux level of the thermal-like component 
reaches as high as $10^4$ keV~ cm$^{-2}$~s$^{-1}$ at high peak energy
($E_{\rm pk}$) ($\sim 300 \pm 100$ keV) even though it is a distant GRB with
$z=1.822$ \citep{Ryde10}.
If we put our simulated GRB at $z=1.822$, the flux level
would be also about $10^4$ keV~cm$^{-2}$~s$^{-1}$ but with lower peak
energy $E_{\rm pk} \sim 190$ keV. 
To explain the discrepancy between the peak energy of
GRB090902B and our model by $\delta(\theta) T$ (Eq.(2)),
this factor should be greater by a factor of 1.57.
On the other hand, since the flux level is comparable between
GRB090902B and our model, it is suggested that the value for
$\delta(\theta)^4 T^4 dS$ is comparable (Eq.(2)).
Thus we can deduce that 
$(\Gamma^2 T'^4 r^2)_{\rm GRB}/(\Gamma^2 
T'^4 r^2)_{\rm Model}$
will be of the order of unity. 
Here $\Gamma$ is the bulk Lorentz factor of the photosphere
and $r$ is the radius of the photosphere.
This is because $\delta(\theta)^4 T^4 dS$ can be rewritten 
as 
$2 \pi \delta(\theta)^4 T^4 r^2 \Gamma^{-2}$ 
(beaming factor), and 
$\delta(\theta) \sim \Gamma$ for a face-on observer. 
Since $r \sim 2\times 10^{12}$ cm and $\Gamma \sim 2\times 10^2$ in our model,
we can constrain the value of ($r/\Gamma$) for GRB090902B as
$4.1 \times 10^{9}$. 
This can be written as $\Gamma \sim 2.4 \times 10^2
(r/ 10^{12} \rm cm )$ in GRB090902B. 
It is noted that the radius of the photosphere depends
on many factors in a complicated way such as the power,
initial Lorentz factor, and mass loading of the jet. 
In this analysis, the radius of the photosphere 
of GRB090902B is assumed to be of the order of $10^{12}$ cm like
the simulation in this study.

There are many explanations for 
many GRBs not showing a  clear thermal spectrum.
The power-law component probably dominates the thermal component,
for example \citet{Peer10}.
For all viewing angles
the derived spectrum below the peak energy is power law with the indices
$2.3 \sim 2.6$.
Maybe this thermal component changes to a non-thermal one
due to the scatterings with non-thermal, high-energy electrons.
GRB jets might be usually dominated by the magnetic field component,
suppressing photospheric emission \citep{zhang09,Zhang11}.
The strict requirement for collimation may also be the reason.

\begin{figure}
\begin{center}
{
\includegraphics[angle=270.,scale=0.33]{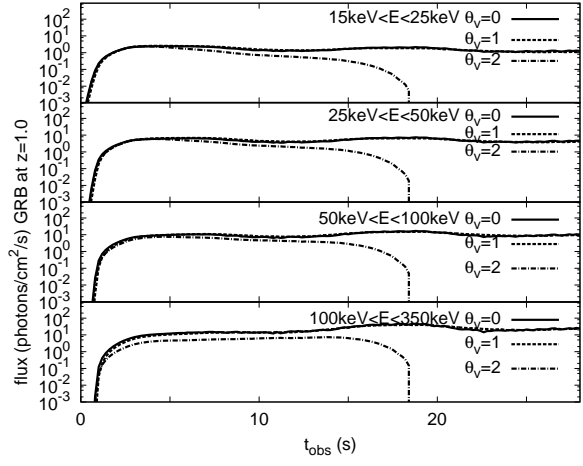}
\caption{
Photon number flux as a function of observer time
for each energy bands (from top to
bottom, $15\rm{keV}<E<25\rm{keV}$,
  $25\rm{keV}<E< 50\rm{keV}$, $50\rm{keV}<E<100\rm{keV}$, $100\rm{keV}<E<350\rm{keV}$).
It is assumed that the burst occurs at the redshift of $z=1$.
The cases of three different viewing angles $\theta_{\rm v}=0^{\circ},~1^{\circ},$
and $2^{\circ}$ are shown.
\label{flux}}}
\end{center}
\end{figure}

\begin{figure}
\begin{center}
{
\includegraphics[angle=270.,scale=0.33]{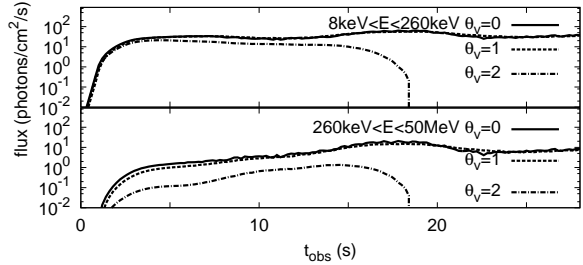}
\caption{Same as Fig. \ref{flux}, but different energy bands 
$8\rm{keV}<E<250\rm{keV}$ (top) and $250\rm{keV}<E<50\rm{MeV}$ (bottom).}
\label{flux2}}
\end{center}
\end{figure}

\begin{figure}
\begin{center}
{
\includegraphics[angle=270.,scale=0.5]{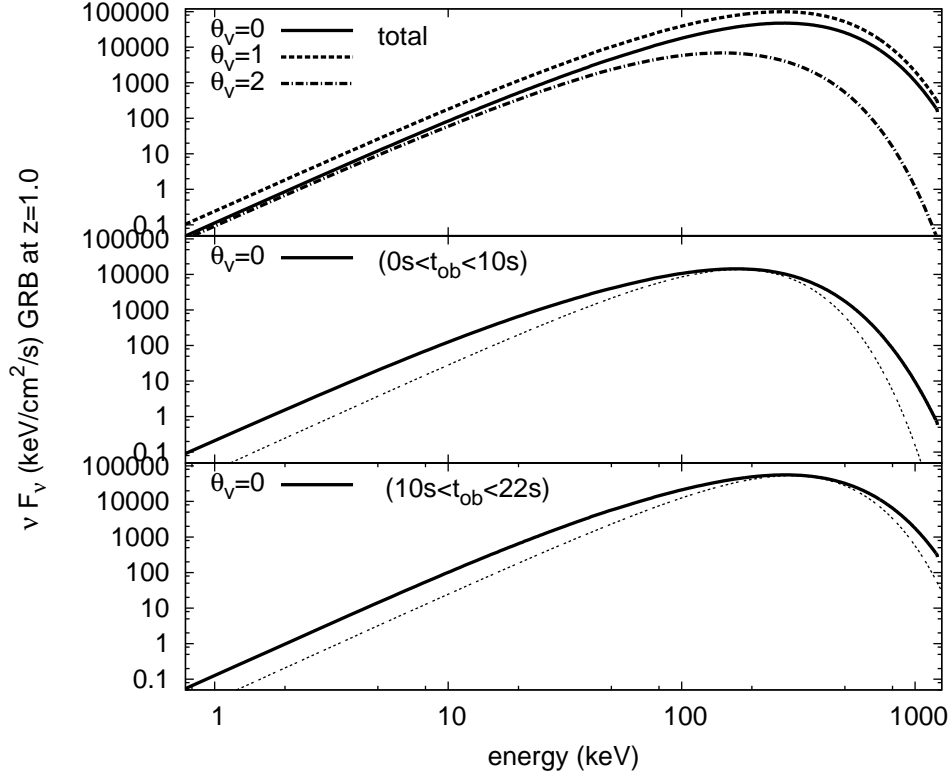}
\vspace{0.5cm}\caption{
$\nu F_{\nu}$ plot for different viewing
angles $\theta_{\rm v}=0^{\circ}, 1^{\circ},$ and $2^{\circ}$.
Total time averaged spectrum are shown (top), and time averaged
plots for two intervals for the on-axis observer are shown (middle and bottom) .
It is assumed that the GRB occurs at the redshift of $z=1$.
Thin  dashed lines shown in middle and bottom panels are
single temperature Plankian distributions
(peak energy and absolute value are fitted.)
The power law indices below the peak energy are $2.3 \sim 2.6$.
\label{nufnu}}}
\end{center}
\end{figure}

The time delay of hard photons shown in Figure \ref{flux} and
\ref{flux2} is of interest, because it may be related to the delayed 
onset observed in some bursts (e.g. GRB080916C, Abdo et al. (2009)).
It should be noted that very high-energy emission may not always correlate 
with lower energy emission, depending how it is created
\citep{peer06,giannios08,lazzati10b}.

\acknowledgments

We thank Hirotaka Itoh and Hiroki Nagakura for useful discussions.
We appreciate Katsuaki Asano for useful comments.
We are grateful to an anonymous for constructive and insightful suggestions.
This work is partly supported by the Grants-in-Aid of the Ministry
of Education, Culture, Sports, and Science and Technology (MEXT) 
(19540236, 21018002 A.M.) and (19047004, 21105509 S.N.), Japan
Society for the Promotion of Science (JSPS) (19104006, 19740139,
21540304 S.N.) and Grant-in-Aid for the Global COE Program 'The Next
Generation of Physics, Spun from Universality and Emergence' from MEXT
of Japan. J.A. is supported by Grant-in-Aid for JSPS Fellows.  
This work was carried out on 
NEC SX8, at
YITP, Kyoto University,
on the Space Science Simulator (NEC SX9) at
JAXA,
and on XT4 at CFCA at NAOJ.

\clearpage

\clearpage

\end{document}